# PERSISTENT CURRENTS IN SUPERCONDUCTING QUANTUM INTERFERENCE DEVICES

F. Romeo

*Dipartimento di Fisica "E. R. Caianiello", Università degli Studi di Salerno*

*I-84081 Baronissi (SA), Italy*

R. De Luca

*CNR-INFM and DIIMA, Università degli Studi di Salerno*

*I-84084 Fisciano (SA), Italy*

## ABSTRACT

Starting from the reduced dynamical model of a two-junction quantum interference device, a quantum analog of the system has been exhibited, in order to extend the well known properties of this device to the quantum regime. By finding eigenvalues of the corresponding Hamiltonian operator, the persistent currents flowing in the ring have been obtained. The resulting quantum analog of the overdamped two-junction quantum interference device can be seen as a supercurrent qubit operating in the limit of negligible capacitance and finite inductance.

PACS: 74.50.+r, 85.25.Dq

Keywords: Josephson junctions, d. c. SQUID, Qubit

# I INTRODUCTION

The d. c. SQUID (Superconducting QUantum Interference Device) is a well known system, widely investigated in the literature [1-3]. This system, though not confined to atomic scale in its dimensions, has been proposed as the basic unit for quantum computing (qubit) by resorting to a characteristic feature of superconductivity: macroscopic quantum coherence [4]. In general, a qubit can be realized by means of a two-level quantum mechanical system [5]. Therefore, the quantum states of a qubit can be a linear combinations of the orthogonal basis $|0\rangle$ and $|1\rangle$, so that the Hilbert space generated by this basis is two-dimensional. Alternatively, a qubit state can be represented by elements of an infinite-dimensional Hilbert space. In this case, however, the effective potential of the system must show a double-well potential, in such a way that one of the two stationary states can be defined as state $|0\rangle$ and the other as state $|1\rangle$.

The electrodynamic properties of d. c. SQUIDs can be analyzed by means of two-junction quantum interferometer models, where each Josephson junction is assumed to be in the overdamped regime. The simplest possible analysis of these systems is done by assuming negligible values of the inductance $L$ of a single branch of the device, so that $\beta = \dfrac{L I_J}{\Phi_0} \approx 0$, where $\Phi_0$ is the elementary flux quantum, and $I_J = \dfrac{I_{J1} + I_{J2}}{2}$ is the mean value of the maximum Josephson currents of the junctions. In this case, the dynamical equation for the superconducting phase differences $\phi_1$ and $\phi_2$ across the two junctions can be written as a single equation for the average phase variable $\varphi = \dfrac{\phi_1 + \phi_2}{2}$. This equation is similar to the nonlinear differential equation governing the time evolution of a single overdamped junction, so that it can be defined as an equivalent single junction model, and is written as follows:

$$\frac{d\varphi}{d\tau} + \cos \pi \psi_{ex} \sin \varphi = \frac{i_B}{2}, \qquad (1)$$

where $\tau = \dfrac{2\pi R I_J}{\Phi_0} t$, with $R = \dfrac{R_1 + R_2}{2}$, $R_1$ and $R_2$ being the resistive junction parameters, $\psi_{ex}$ is the externally applied flux normalized to $\Phi_0$ and $i_B$ is the bias current normalized to $I_J$.

Following the same type of approach, by means of a perturbation analysis, taking $\beta$ as the perturbation parameter, it can be shown that, to first order in $\beta$, the equivalent single junction model can be written as follows for a symmetric SQUID with identical junctions [6]:

$$\frac{d\varphi}{d\tau} + (-1)^n \cos(\pi \Psi_{ex}) \sin \varphi + \pi \beta \sin^2(\pi \Psi_{ex}) \sin 2\varphi = \frac{i_B}{2}, \qquad (2)$$

where $n$ is an integer. This model allows, at least for small values of the parameter, to calculate in closed form some electrodynamic quantities, such as, for example, the amplitude of the half-integer Shapiro steps appearing in these systems [7]. It has also been shown that, by extending this model to SQUIDs with non-identical junction, one can obtain an effective classical double-well potential in which the transition from one state to the other can be enhanced by applying an opportune external magnetic flux [8]. However, this classical analysis by itself does not allow to define the quantum states of the system. Nonetheless, the equation of the motion (2) could be assumed to be a classical version of the time evolution of a quantum phase state. Therefore, the aim of the present work is to obtain, starting from the time evolution of the superconducting phase difference $\varphi$, the quantum mechanical Hamiltonian and to compute, by means of this quantum mechanical system, which in the classical limit reduces to Eq. (2), the persistent currents in the SQUID. It is interesting to notice that the resulting "Hamiltonian" quantum model derives from the overdamped limit of a classical dissipative system in the presence of a double well potential. The present analysis can be seen as an alternative approach to the study of the quantum properties of supercurrent qubits: It allows to study the response of the quantum system in the limit of negligible capacitance and finite values of the inductance, as opposed to the case usually considered in the literature, where negligible inductance and finite capacitance is assumed [9 - 10].

## II FROM CLASSICAL TO QUANTUM MECHANICS

Let us consider the classical dynamical equation $\dot{x} = f(x)$ for the state variable $x$, where the dot notation indicates the derivative with respect to the normalized time $\tau$. Making use of the previous equation, taking the time derivative of both sides, we can obtain the equation of the motion of the quantity $\dot{x}$ as follows:

$$\ddot{x} = f_x(x)\dot{x} = f_x(x)f(x),\qquad(3)$$

where the notation $f_x(x)$ stands for the partial derivative of $f(x)$ with respect to $x$. Given the above equation and following the procedure described by Huang and Lin [11], the Lagrangian associated to this problem is obtained in the form:

$$L = \frac{1}{2}\left[\dot{x}^2 + (f(x))^2\right].\qquad(4)$$

Starting from the Lagrangian $L$, the Legendre transformation allows us to get the following classical Hamiltonian:

$$H = \frac{1}{2}\left[\pi^2 - (f(x))^2\right],\qquad(5)$$

where $\pi = \dfrac{\partial L}{\partial \dot{x}}$ is the canonical momentum conjugated to the variable $x$, while $U(x) = -\dfrac{1}{2}(f(x))^2$ could be considered as an effective potential. We are now interested in the quantization of the classical model described so far. According to the standard procedure, the recipe to transform the classical Hamiltonian in a quantum operator is implemented by making the substitution $\pi \to \hat{\pi} = -i\partial_x, \quad x \to \hat{x} = x$ (in dimensionless units). From the previous definitions, the commutation rule $\left[\hat{x}, \hat{\pi}\right] = i$ for the conjugated variables follows directly. Furthermore, the Hamiltonian operator can be written as $\hat{H} = -\dfrac{1}{2}\left[\partial_x^2 + (f(x))^2\right]$.

The general procedure described above can be adopted to obtain the quantum model of an overdamped d.c. SQUID in the limit in which the reduced two-junctions interferometer model [6] can be applied. In the framework of this model, the phase dynamics can be written (in the homogeneous case) as in Eq. (2), so that the function $f = f(\varphi)$ takes the following form:

$$f(\varphi) = \gamma - a \sin(\varphi) - b \sin(2\varphi), \qquad (6)$$

where $a = \cos(\pi \psi_{ex})$, $b = \pi \beta \sin^2(\pi \psi_{ex})$, $\gamma = \dfrac{i_B}{2}$, having chosen $n = 0$ for simplicity. We notice that this analysis cannot be extended to the similar case, considered by Grønbech-Jensen et al. [12], of junctions with finite capacitance. Therefore, by setting $x = \varphi$ and $\pi = \dot{\varphi}$ in the above general analysis, we notice that the phase and the voltage across the two-junction quantum interferometer are conjugate variables of the system. In the present case, therefore, proceeding as we said, by squaring $f(\varphi)$ and exhibiting the final result of the calculation in terms of the higher harmonics of the phase variable instead of powers of trigonometric functions, the following Hamiltonian operator is obtained:

$$\hat{H} = \frac{-\partial_{\varphi}^2}{2} - \frac{ab}{2}\cos(\varphi) + \frac{a^2}{4}\cos(2\varphi) + \frac{ab}{2}\cos(3\varphi) + \frac{b^2}{4}\cos(4\varphi),$$
$$+ \gamma\, a \sin(\varphi) + \gamma\, b \sin(2\varphi) - \Xi(a,b,\gamma) \qquad (7)$$

where $\Xi(a,b,\gamma) = \dfrac{2\gamma^2 + a^2 + b^2}{4}$ is a flux dependent energy shift which will be important in the following discussion.

In order to calculate the relevant physical quantities of the system, we introduce the orthonormal complete basis $\langle \varphi | n \rangle = \dfrac{e^{in\varphi}}{\sqrt{2\pi}}$ of the infinite-dimensional Hilbert space with the inner product $\langle m | n \rangle = \displaystyle\int_0^{2\pi} \frac{d\varphi}{2\pi} e^{i(n-m)\varphi} = \delta_{n,m}$, where $\delta_{n,m}$ is the Kronecker delta. In this representation the matrix elements of the Hamiltonian operator can be written as follows:

$$H_{n,m} = \frac{n^2 - 2\,\Xi(a,b,\gamma)}{2}\delta_{n,m} - \frac{ab}{4}\left(\delta_{n,m-1} + \delta_{n,m+1}\right) + \frac{a^2}{8}\left(\delta_{n,m-2} + \delta_{n,m+2}\right) +$$
$$+ \frac{ab}{4}\left(\delta_{n,m-3} + \delta_{n,m+3}\right) + \frac{b^2}{8}\left(\delta_{n,m-4} + \delta_{n,m+4}\right) + \frac{a\gamma}{2i}\left(\delta_{n,m-1} - \delta_{n,m+1}\right) + \frac{b\gamma}{2i}\left(\delta_{n,m-2} - \delta_{n,m+2}\right) \qquad (8)$$

where the following useful relations have been used:

$$\langle m|\sin(l\varphi)|n\rangle = \frac{1}{2i}\left(\delta_{n,m-l} - \delta_{n,m+l}\right), \tag{9a}$$

$$\langle m|\cos(l\varphi)|n\rangle = \frac{1}{2}\left(\delta_{n,m-l} + \delta_{n,m+l}\right). \tag{9b}$$

Once the matrix elements of the Hamiltonian operator are known, we can diagonalize a reduced version of the complete infinite-dimensional matrix by introducing an energy cut-off. Such procedure can be safely carried out when we need to characterize low energy states which are located very far from the cut and when the number of the vectors in the basis of the reduced Hilbert space is able to capture the essential features of the low energy states. For instance, the Hilbert space spanned by the first 20 basis functions can be a very effective choice, if we need to study only the lowest energy states close to the ground state of our system. In fact, in our case we have noted that, by halving the number of the basis elements, no evident difference is present in the lowest energy eigenvalues.

## III  PERSISTENT CURRENTS

Following the procedure described above, in the present section we shall derive the behavior of the persistent currents associated to each eigenstate of the Hamiltonian as a consequence of the time reversal symmetry breaking provided by the magnetic flux. Such a current, in units of the Josephson current divided by $2\pi$ (i. e., in units of $\frac{E_J}{\Phi_0}$, where $E_J$ is the Josephson energy), can be defined as follows:

$$I_n = -\frac{\partial \varepsilon_n}{\partial \psi_{ex}}, \tag{10}$$

where $\varepsilon_n$ and $\psi_{ex}$ are the eigenvalues of the Hamiltonian and the normalized external magnetic flux, respectively. According to the above relation, the persistent current $I_n$ can be computed once the pertinent eigenvalue $\varepsilon_n$ is known. Furthermore, it should be noticed that, in the absence of the off-diagonal terms in the Hamiltonian given in Eq. (8), the state independent persistent current computed by means of Eq. (10) would be given by:

$$I = -\frac{\pi}{4}\sin(2\pi\psi_{ex})\left[1 - 2\pi^2\beta^2\sin^2(\pi\psi_{ex})\right]. \tag{11}$$

The solution of the full problem can thus be seen as the state dependent correction to the above relation induced by the off-diagonal terms. Last point can be well understood by analyzing Figs. 1a-b. In these figures, even tough we are in the presence of finite off-diagonal terms, the relation given in Eq. (11) is able to describe quite accurately the behavior of the persistent currents which appears insensitive to the state index due to the small value of $\beta$. When the value of $\beta$ is raised (see Fig. 2a), the persistent currents starts to become weakly state sensitive and some deformation of the original shape occurs. Furthermore, the states of higher energy (see Fig. 2b) induce a behavior of the persistent current which is quite insensitive to the state index. A further raising of $\beta$ (see Fig. 3a) induces a suppression of the persistent current carried by the first excited state in the vicinity of half integer values of the normalized applied magnetic flux. This implies that, in the low energy regime (i. e., when the quantum state can be written as $|S\rangle = |\widetilde{0}\rangle\langle\widetilde{0}|S\rangle + |\widetilde{1}\rangle\langle\widetilde{1}|S\rangle$, where $|\widetilde{0}\rangle$ and $|\widetilde{1}\rangle$ represent the ground state and the first excited state, respectively), the average persistent current $\langle I\rangle = \left|\langle\widetilde{0}|S\rangle\right|^2 I_0 + \left|\langle\widetilde{1}|S\rangle\right|^2 I_1$ close to an half integer flux is mainly related to the ground state properties of the system, since $I_0 >> I_1$ in the vicinity of $\psi_{ex} = \frac{1}{2}$ (for $\psi_{ex} \neq \frac{1}{2}$). In Fig. 4a, raising once again $\beta$, it can be noticed that, in the vicinity of half integer values of the normalized flux, the ground state and the first excited state carries currents of opposite sign, inducing a competing magnetic behavior. Therefore, the average persistent current, and its magnetic behavior, depend, on both coefficients of the decomposition (i. e., on $\langle\widetilde{0}|S\rangle$ and $\langle\widetilde{1}|S\rangle$). This last point implies that, by measuring the magnetic momentum of the system in a particular magnetic field configuration, we can obtain constraints on the nature of the quantum superposition. For instance, under these conditions, we could prepare the quantum state in such a way that the average persistent current is negligibly small in the vicinity of half integer values of the normalized applied magnetic flux.

Furthermore, we point out that a double well potential can be obtained setting the model parameters as done in Fig.5 ( $\beta = 0.2$ and $\psi_{ex} = 0.7$ ), where the potential $U(\varphi)$ is shown. Indeed, we notice that for $\gamma = 0$ two low-energy degenerate states are present, the degeneracy being removed by means of a small current bias. Such bias can drive the response of the system toward one of the two minima of the potential allowing a complete control of the quantum state which can be exploited for technological applications. Finally, we notice that, even thought the chosen $\beta$ values in Figs. 2 − 5 are close to the validity limits of the first order approximation of the reduced model in ref. [6], the above characteristic response of the system remain qualitatively valid, since we are here considering the leading order in the value of $\beta$ .

## IV  CONCLUSION

Starting from the reduced dynamical model of the two-junction quantum interference device, the applied flux dependence of persistent currents in this system has been studied in the quantum regime. The extension of the dissipative overdamped classical system, from classical to quantum mechanics, allows to consider the electrodynamical response of a supercurrent quantum bit in the limit of negligible capacitance and finite inductance. For null bias current and for opportune values of the externally applied magnetic flux, the quantum analog of the two-junction interferometer shows effective potential with a degenerate ground state; degeneracy can be removed by applying a control non-null bias current.

In the literature, the quantum behavior of the two-junction quantum interference device is studied by considering the charging energy of the junctions as preponderant with respect to the energy of the circulating currents [5, 9, 10]. In the present work it is shown that it is possible to obtain an Hamiltonian quantum analog of d. c. SQUIDs containing overdamped junctions in the limit of null capacitance and finite inductance values. In this framework, a flux qubit can be realized, under quite different conditions than those with high junction capacitance value [5]. Finally, the present analysis can also be considered as a link between classical dissipative systems and their corresponding quantum mechanical models.

## FIGURE CAPTIONS

### Fig. 1

**(a)** Persistent currents $I_1$ (triangle) and $I_2$ (box) plotted as a function of the applied external flux $\psi_{ex}$ and by fixing $\beta = 0.075$ and $\gamma = 0$. **(b)** Persistent currents $I_3$ (star) and $I_4$ (diamond) plotted as a function of the applied external flux $\psi_{ex}$ and by fixing $\beta = 0.075$ and $\gamma = 0$.

### Fig. 2

**(a)** Persistent currents $I_1$ (triangle) and $I_2$ (box) plotted as a function of the applied external flux $\psi_{ex}$ and by fixing $\beta = 0.15$ and $\gamma = 0$. **(b)** Persistent currents $I_3$ (star) and $I_4$ (diamond) plotted as a function of the applied external flux $\psi_{ex}$ and by fixing $\beta = 0.15$ and $\gamma = 0$.

### Fig. 3

**(a)** Persistent currents $I_1$ (triangle) and $I_2$ (box) plotted as a function of the applied external flux $\psi_{ex}$ and by fixing $\beta = 0.2$ and $\gamma = 0$. **(b)** Persistent currents $I_3$ (star) and $I_4$ (diamond) plotted as a function of the applied external flux $\psi_{ex}$ and by fixing $\beta = 0.2$ and $\gamma = 0$.

### Fig. 4

**(a)** Persistent currents $I_1$ (triangle) and $I_2$ (box) plotted as a function of the applied external flux $\psi_{ex}$ and by fixing $\beta = 0.25$ and $\gamma = 0$. **(b)** Persistent currents $I_3$ (star) and $I_4$ (diamond) plotted as a function of the applied external flux $\psi_{ex}$ and by fixing $\beta = 0.25$ and $\gamma = 0$.

### Fig. 5

Density plot of the potential $U(\varphi)$ plotted as a function of the phase $\varphi$ and of the normalized bias current $\gamma$ by setting the remaining parameters as: $\beta = 0.2$ and $\psi_{ex} = 0.7$. Lower energy states are represented by darker regions in the plot.

**Fig. 1**

**(a)**

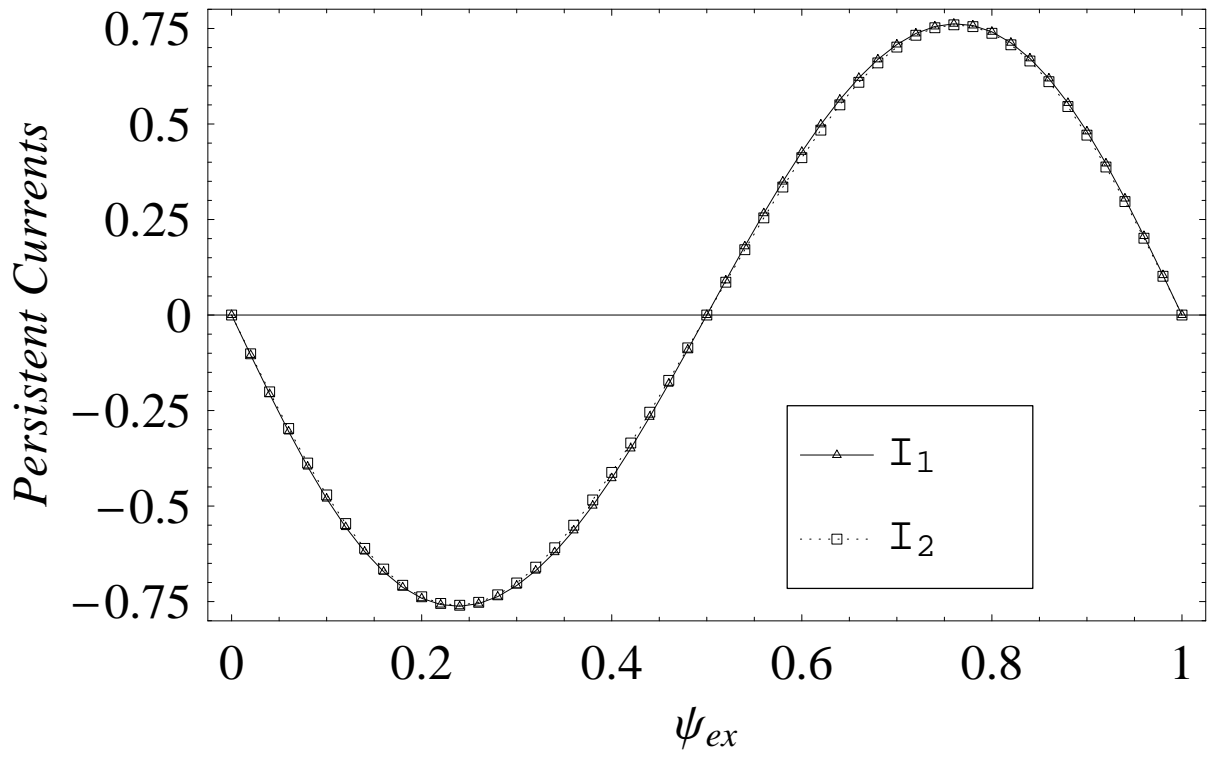

**(b)**

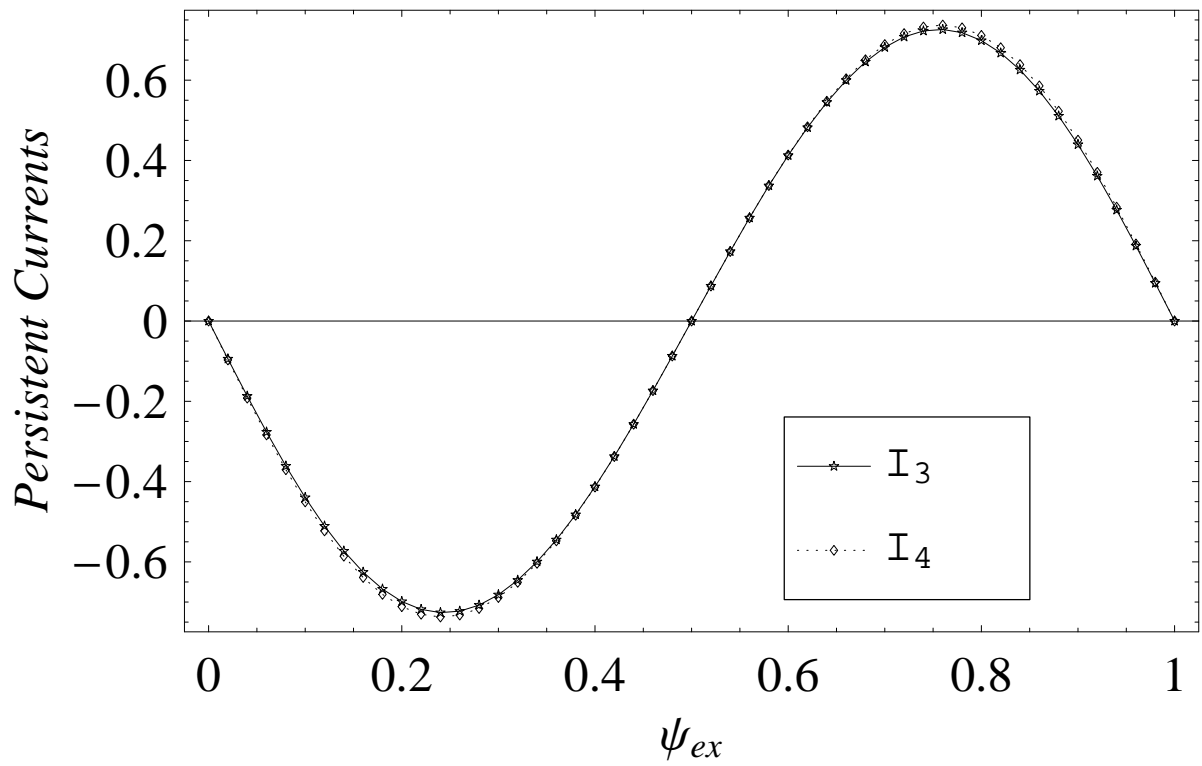

**Fig. 2**

**(a)**

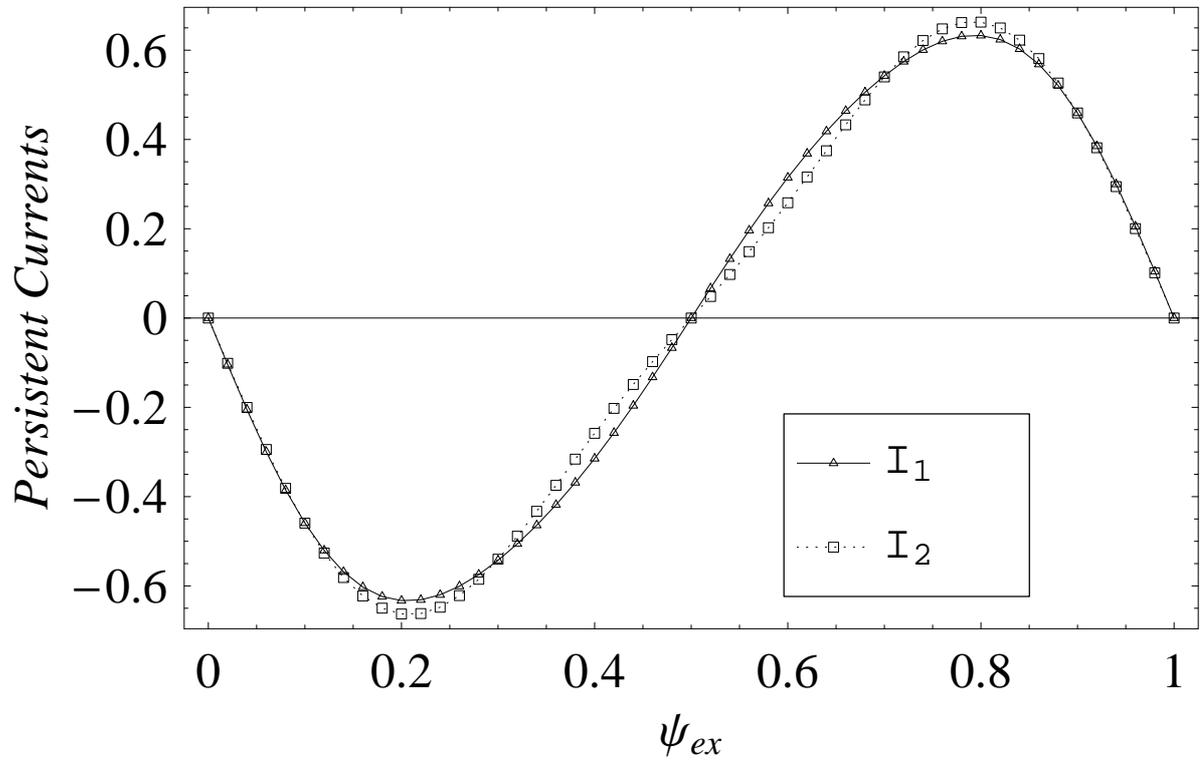

**(b)**

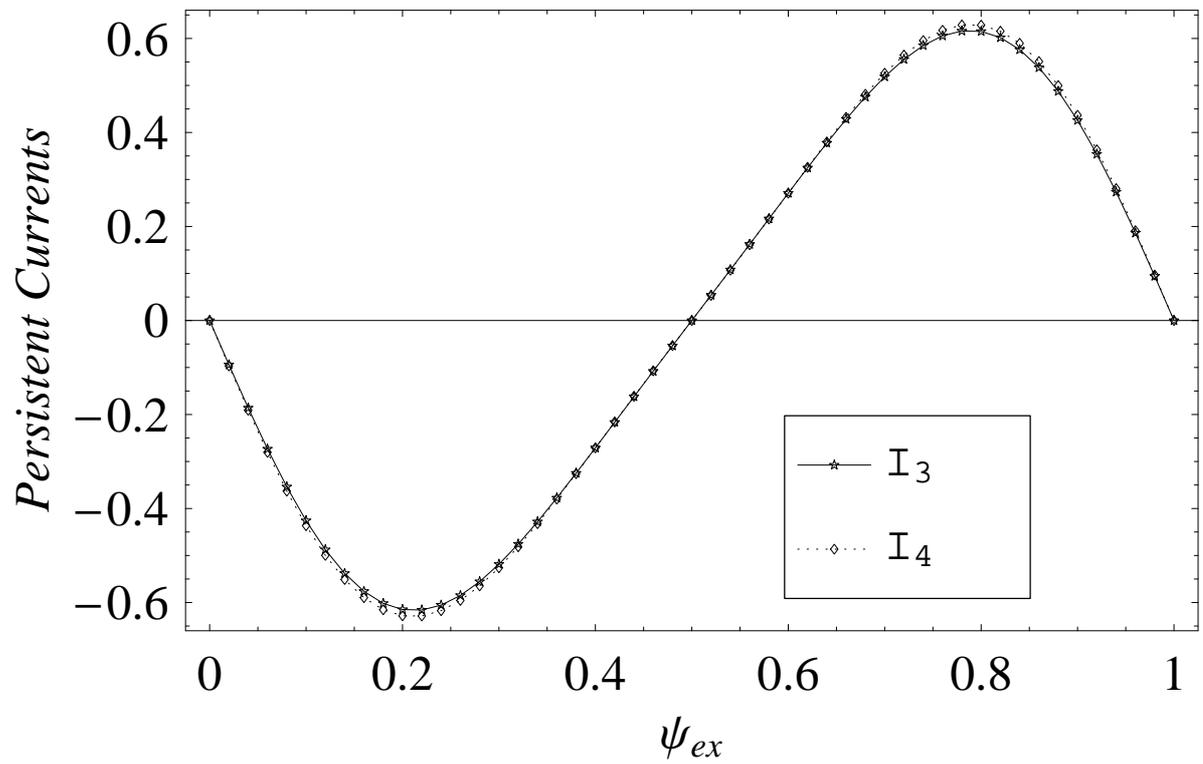

**Fig. 3**

**(a)**

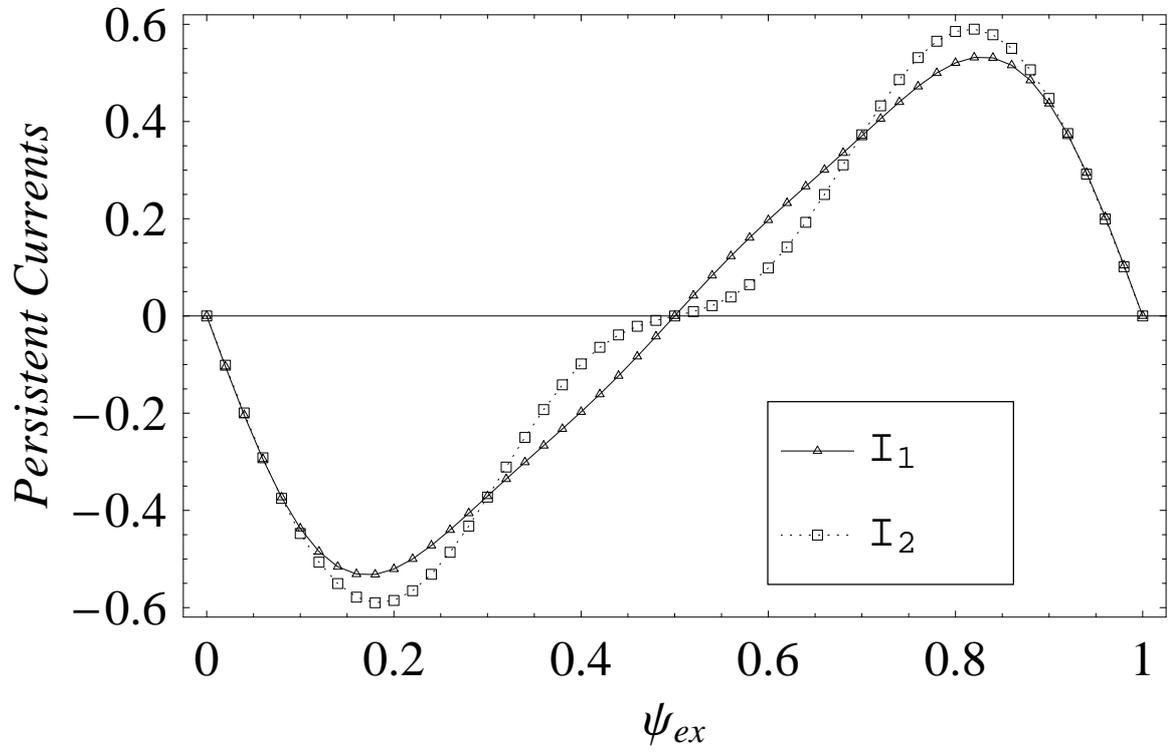

**(b)**

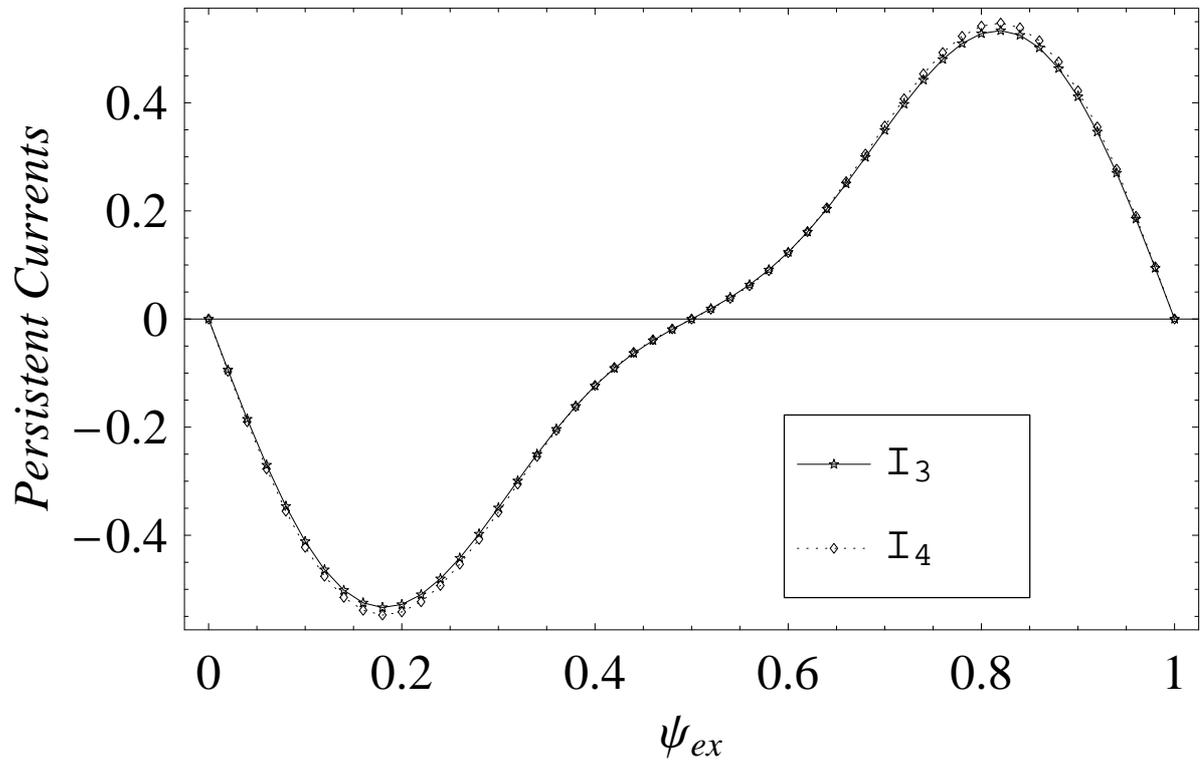

**Fig. 4**

**(a)**

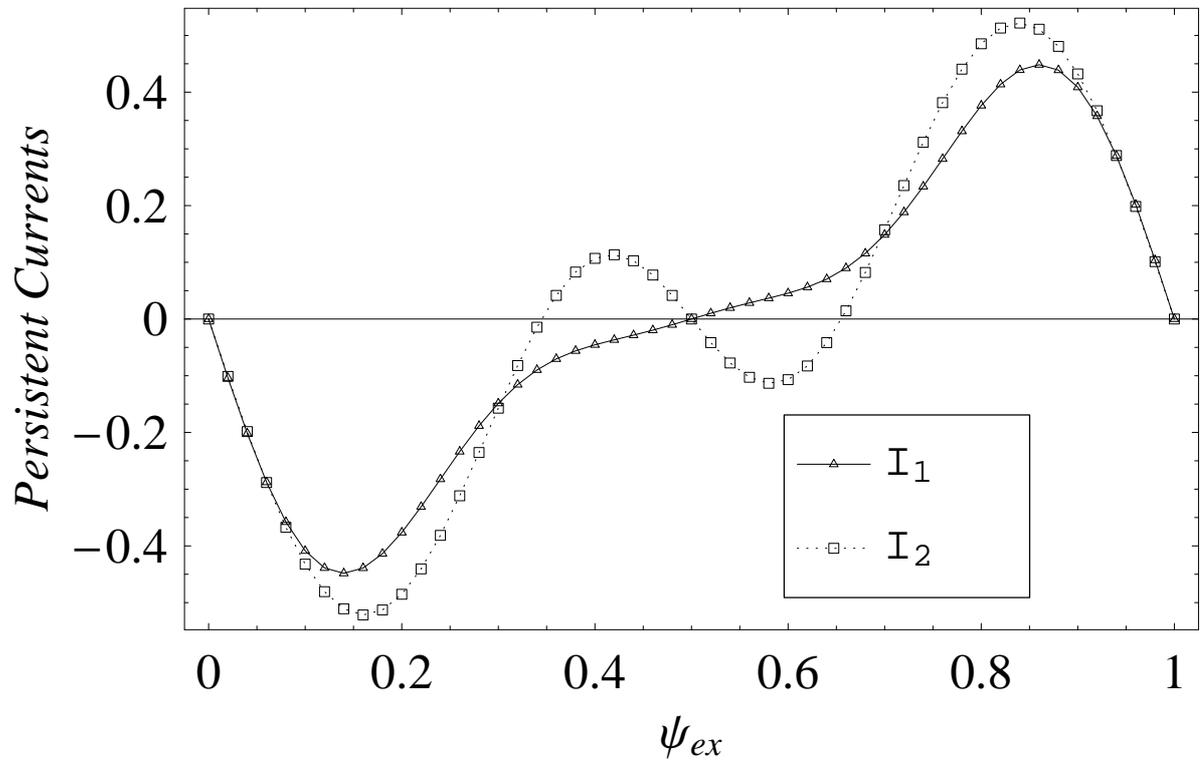

**(b)**

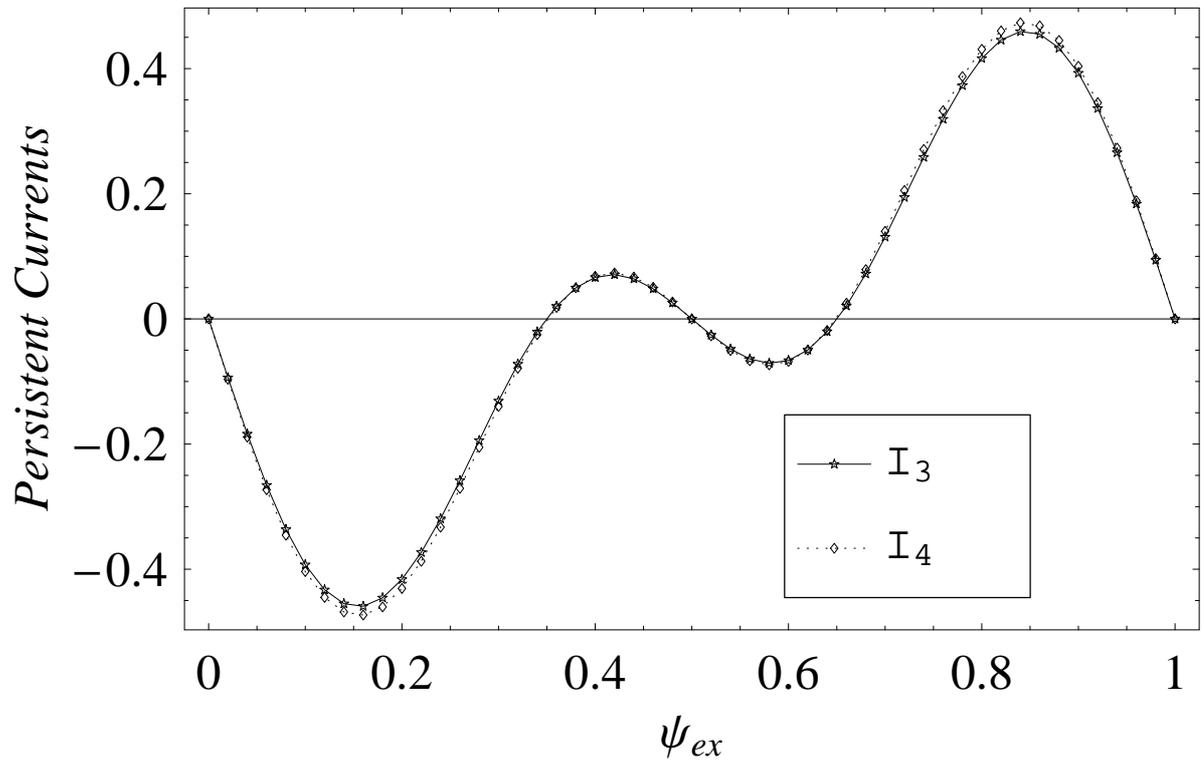

**Fig. 5**

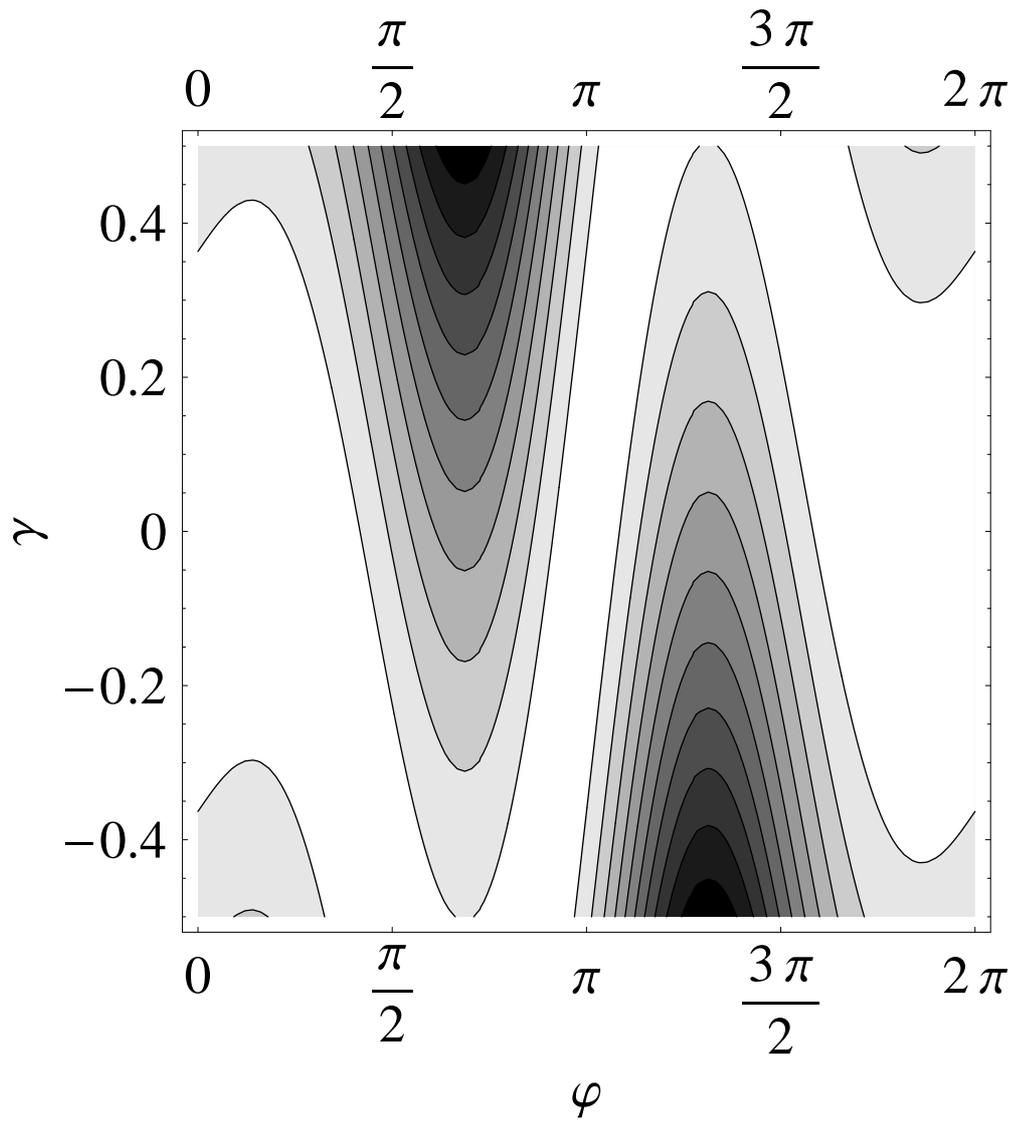